\documentclass[9pt,twocolumn,twoside]{pnas-new}
\usepackage{todonotes}
\templatetype{pnasresearcharticle} 

\title{Visualizing probabilistic models: Intensive~Principal~Component~Analysis}

\author[a,1]{Katherine N. Quinn}
\author[a]{Colin B. Clement} 
\author[a]{Francesco De Bernardis}
\author[a]{Michael D. Niemack}
\author[a]{James P. Sethna}

\affil[a]{Department of Physics, Cornell University, Ithaca, NY 14853-2501, United States}

\leadauthor{Quinn} 

\significancestatement{We introduce Intensive Principal Component Analysis (InPCA), a widely applicable manifold-learning method to visualize general probabilistic models and data. Using replicas to tune dimensionality in high-dimensional data, we use the zero-replica limit to discover a distance metric which preserves distinguishability in high dimensions, and an embedding with superior visualization performance. We apply InPCA to the model of cosmology which predicts the angular power spectrum of the cosmic microwave background, allowing visualization of the space of model predictions (i.e. different universes).}

\authorcontributions{KNQ, CBC, and JPS designed research; KNQ and CBC performed research; FDB and MDN provided expertise on CMB; KNQ, CBC, FDB, MDN, and JPS wrote the paper.}
\correspondingauthor{\textsuperscript{1}To whom correspondence should be addressed. E-mail: knq2\@cornell.edu}

\keywords{Manifold Learning $|$ Information Theory $|$ Statistical Physics $|$ Probabilistic Models $|$ Probabilistic Data $|$ Visualization} 

\begin{abstract}

Unsupervised learning makes manifest the underlying structure of data without curated training and specific problem definitions. However, the inference of relationships between data points is frustrated by the `curse of dimensionality' in high-dimensions. Inspired by replica theory from statistical mechanics, we consider replicas of the system to tune the dimensionality and take the limit as the number of replicas goes to zero. The result is the intensive embedding, which is not only isometric (preserving local distances) but allows global structure to be more transparently visualized. We develop the Intensive Principal Component Analysis (InPCA) and demonstrate clear improvements in visualizations of the Ising model of magnetic spins, a neural network, and the dark energy cold dark matter (\LCDM) model as applied to the Cosmic Microwave Background.

\end{abstract}

\dates{This manuscript was compiled on \today}
\doi{\url{www.pnas.org/cgi/doi/10.1073/pnas.XXXXXXXXXX}}

\graphicspath{{images/}}

\newcommand{\B}[1]{{\boldsymbol{#1}}}
\newcommand{\Like}[1]{{\mathcal{L}\left(#1\right)}}

\newcommand{\LCDM}{$\Lambda$CDM}
\newcommand{\norm}[1]{\left\lVert#1\right\rVert}

\begin{document}

\maketitle
\thispagestyle{firststyle}
\ifthenelse{\boolean{shortarticle}}{\ifthenelse{\boolean{singlecolumn}}{\abscontentformatted}{\abscontent}}{}

\dropcap{V}isualizing high-dimensional data is a cornerstone of machine learning, modeling, big data, and data mining. These fields require learning faithful and interpretable low-dimensional representations of high-dimensional data, and, almost as critically, producing visualizations which allow interpretation and evaluation of what was learned~\cite{Ferreira2003,Liu2017,Lee2007,ArthurZimekErichSchubert2012}.
Unsupervised learning, which infers features from data without manually curated data or specific problem definitions~\cite{murphy2012machine}, is especially important for high-dimensional, big data applications in which specific models are unknown or impractical. For high dimensions, the relative distances between features become small and most points are orthogonal to one another~\cite{kriegel2009clustering}. A trade-off between preserving local and global structure must often be made when inferring a low-dimensional representation. Classic manifold learning techniques include linear methods such as Principal Component Analysis (PCA)~\cite{hotelling1933} and multidimensional scaling (MDS)~\cite{Torgerson1952}, which preserve global structure but at the cost of obscuring local features. Existing nonlinear manifold learning techniques, such as t-distributed Stochastic Network Embedding (t-SNE)~\cite{VanDerMaaten2008} and diffusion maps~\cite{Coifman2005}, preserve the local structure while only maintaining some qualitative global patterns such as large clusters. The Uniform Manifold Approximation (UMAP)~\cite{mcinnes2018umap} better preserves topological structures in data, a global property. 

In this manuscript, we develop a new nonlinear manifold learning technique which achieves a compromise between preserving local and global structure. We accomplish this by developing an isometric embedding for general probabilistic models based on the replica trick~\cite{MezardPV87}. Taking the number of replicas to zero, we reveal an intensive property~--~an information density characterizing the distinguishability of distributions~--~ameliorating the canonical orthogonality problem and `curse of dimensionality.' We then describe a simple, deterministic algorithm that can be used for any such model, which we call Intensive Principal Component Analysis~(InPCA).  Our method quantitatively captures global structure while preserving local distances. We first apply InPCA to the canonical Ising model of magnetism, which inspired the zero-replica limit. Next, we show how InPCA can capture and summarize the learning trajectory of a neural network. Finally, we visualize the dark energy cold dark matter (\LCDM) model as applied to the Cosmic Microwave Background (CMB), using InPCA, t-SNE and Diffusion Maps.

\section*{Model Manifolds of Probability Distributions}
Any measurement obtained from an experiment with uncertainty can generally be understood as a probability distribution. For example, when some data $x$ is observed with normally distributed noise $\xi$  of variance $\sigma^2$, under experimental conditions $\B{\theta}_j$, a model is expressed as
\begin{eqnarray}
x = f(\B{\theta}_j) + \xi \quad \mathrm{where} \quad \mathcal{L}(\xi) \backsim\ \mathcal{N}(0,\sigma^2),
\end{eqnarray}
and $f(\B{\theta}_i)$ is a prediction given the experimental conditions. This relationship is equivalent to saying that the probability of measuring data $x$ given some conditions $\B{\theta}$ is:
\begin{equation}
    \mathcal{L}(x \mid \B{\theta} ) \backsim\ \mathcal{N}(f(\B{\theta}),\sigma^2).
\end{equation}
More complicated noise profiles with asymmetry or correlations can be accommodated with this picture. Measurements without an underlying model can also be seen as distributions, where a measurement $x_i$ with uncertainty $\sigma$ can induce a probability $\Like{x\mid x_i,\sigma}$ of observing new data $x$.

We define a probabilistic model $\Like{\B{x}\mid\B{\theta}}$, the likelihood of observing data $\B{x}$ given parameters $\B{\theta}$. The \textit{model manifold} is defined as the set of all possible predictions, $\{\Like{\B{x}\mid\B{\theta}_i}\}$, which is a surface parameterized by the model parameters $\{\B{\theta}_i\}$. The parameter directions related to the longest distances along the model manifold have been shown to predict emergent behavior (how microscopic parameters lead to macroscopic behavior)~\cite{MachtaCTS13}. We will see than InPCA orders its principal components by the length of the model manifold along their direction, highlighting global structure. The boundaries of the model manifold represent simplified models which retain predictive power~\cite{Transtrum2014}, and the constraint of data lying near the model manifold has been used to optimize experimental design~\cite{TranstrumSloppyReview}. In this manuscript, we study the Ising model, which defines probabilities of spin configurations given interaction strengths, a neural network, which predicts the probability of an image representing a single handwritten digit given weights and biases, and \LCDM, which predicts the distribution of CMB radiation given fundamental constants of nature.

\section*{Hypersphere Embedding}

We promised an embedding which is both isometric, and preserves global structures. We satisfy the first promise by considering the hypersphere embedding:
\begin{equation}
   \{z_\B{x}(\B{\theta_i})\} = \left\{2\sqrt{\Like{\B{x}\mid\B{\theta_i}}}\right\},
   \label{eq:firstembedding}
\end{equation}
where the normalization constraint of $\Like{\B{x}\mid\B{\theta}}$ forces $z_\B{x}$ to lie on the positive orthant of a sphere. A natural measure of distance on the hypersphere is the Euclidean distance, in this case also known as the Hellinger divergence~\cite{hellinger}
\begin{eqnarray}
\label{eq:hellinger}
d^2(\B{\theta_1},\B{\theta_2}) &=& \norm{\B{z}(\B{\theta_1}) - \B{z}(\B{\theta_2})}^2 \nonumber \\
&=& 8\left(1-\sqrt{\Like{\B{x}\mid\B{\theta}_1}}\cdot\sqrt{\Like{\B{x}\mid\B{\theta}_2}}\right)^2,
\label{eq:spheredist}
\end{eqnarray}
where $\cdot$ represents the inner product over $x$. Now we can see that the hypersphere embedding is isometric:
the Euclidean metric of this embedding is equal to the Fisher Information metric $\mathcal{I}$ of the model manifold~\cite{gromov2013search},
\begin{equation}
	d^2(z_i,z_i+dz_i) = \sum_i dz_i dz_i 
    = \sum_{kl} \mathcal{I}_{kl} d\theta_kd\theta_l.
    \label{eq:FIM}
\end{equation}
The Fisher Information Metric (FIM) is the natural metric of the model manifold~\cite{Amari}, so the hypersphere embedding preserves the local structure of the manifold defined by $\Like{\B{x}\mid\B{\theta}}$.

As the dimension of the data increases, almost all features become orthogonal to each other, and most measures of distance lose their ability to discriminate between the smallest and largest distances~\cite{beyer1999nearest}. For the hypersphere embedding, we see that as the dimension of $x$ increases, the inner product in the Hellinger distance of Eq.~\ref{eq:spheredist} becomes smaller as the probability is distributed over more dimensions. In the limit of large dimension, all non-identical pairs of points become orthogonal and equidistant around the hypersphere (a constant distance $\sqrt{8}$ apart), frustrating effective dimensional reductions and visualization.

\begin{figure}

\begingroup%
  \makeatletter%
  \providecommand\color[2][]{%
    \errmessage{(Inkscape) Color is used for the text in Inkscape, but the package 'color.sty' is not loaded}%
    \renewcommand\color[2][]{}%
  }%
  \providecommand\transparent[1]{%
    \errmessage{(Inkscape) Transparency is used (non-zero) for the text in Inkscape, but the package 'transparent.sty' is not loaded}%
    \renewcommand\transparent[1]{}%
  }%
  \providecommand\rotatebox[2]{#2}%
  \ifx\svgwidth\undefined%
    \setlength{\unitlength}{224.36037598bp}%
    \ifx\svgscale\undefined%
      \relax%
    \else%
      \setlength{\unitlength}{\unitlength * \real{\svgscale}}%
    \fi%
  \else%
    \setlength{\unitlength}{\svgwidth}%
  \fi%
  \global\let\svgwidth\undefined%
  \global\let\svgscale\undefined%
  \makeatother%
  \begin{picture}(1,0.97090274)%
    \put(0,0){\includegraphics[width=\unitlength,page=1]{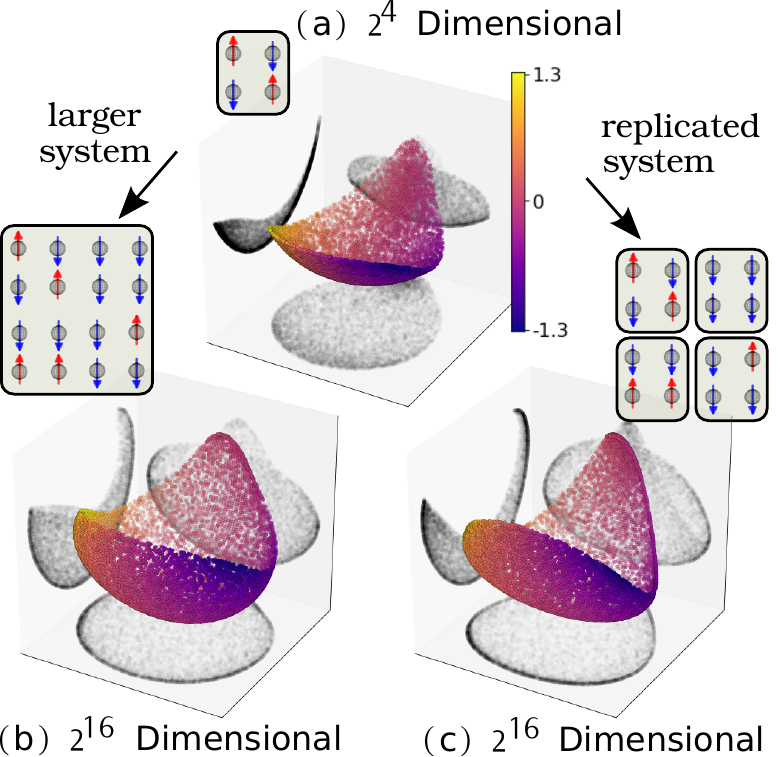}}%
  \end{picture}%
\endgroup%

    \caption{ \textbf{Hypersphere embedding}, illustrating an embedding of the two dimensional Ising model. Points were generated through a Monte Carlo sampling and visualized by projecting the probability distributions onto the first 3 principal components. The points are colored by magnetic field strength. As the system size increases from $2\times 2$ to $4\times 4$, the orthogonality problem is demonstrated by an increase in `wrapping' around the hypersphere. This effect can be also be produced by instead considering four replicas of the original system, motivating the replica trick which takes the embedding dimension or number of replicas to zero.}
    \label{fig:hypersphere}
\end{figure}

To illustrate this problem with the hypersphere embedding, consider the Ising Model, which predicts the likelihood of observing a particular configuration of binary random variables (spins) on a lattice. The probability of a spin configuration is determined by the Boltzmann distribution, and is a function of a local pairwise coupling, and a global applied field. The dimension is determined by the number of spin configurations, $2^N$ where $N$ is the number of spins. Holding temperature fixed at one, we vary $h$ and $J$: external magnetic field ($h\in (-1.3,1.3)$) and nearest neighbour coupling ($J\in (-0.4,0.6)$), using a Monte Carlo method weighted by Jeffrey's Prior to sample 12,000 distinct points. From the resulting set of parameters, we compute \smash{$X_{ij}=\{\B{z}_i(\B{\theta_j})\}$} using the Boltzmann distribution, and visualize the model manifold in the $N$-sphere embedding of Eq.~\ref{eq:firstembedding} by projecting the predictions onto the first three principal components of $X$. Figure~\ref{fig:hypersphere}{(a)} shows this projection of the model manifold of a $2\times 2$ Ising model which is embedded in $2^4$ dimensions. Figure~\ref{fig:hypersphere}{(b)} shows a larger, $4\times 4$ Ising model, of dimension $2^{16}$. As the dimension is increased from $2^4$ to $2^{16}$, we see the points starting to wrap around the hypersphere, becoming increasingly equidistant and less distinguishable.

A natural way to increase the dimensionality of a probabilistic model is by drawing multiple samples from the distribution. If $D$ is the dimension of $x$, then $N$ identical draws from the distribution will have dimension $D^N$. The more samples drawn, the easier it is to distinguish between distributions, mimicking the `curse of dimensionality' for large systems. We see this demonstrated for our Ising model in Fig.~\ref{fig:hypersphere}{(c)}, where we drew 4 replica samples from the same model. Notice that as compared to the original 2$\times$2 model, the model manifold of the 4-replica 2$\times$2 model `wraps' more around the hypersphere, just like the larger, $4\times 4$ Ising model. High dimensional systems have `too much information,' in the same way that large numbers of samples have too much information. In the next section, we consider the contraposition of the insight that a large number of replicas leads to the the curse of dimensionality, and discover an embedding which is not only isometric but also ameliorates the high-dimensional wrapping around the $n$-sphere.

\section*{Replica Theory and the Intensive Embedding}

\begin{figure}

\begingroup%
  \makeatletter%
  \providecommand\color[2][]{%
    \errmessage{(Inkscape) Color is used for the text in Inkscape, but the package 'color.sty' is not loaded}%
    \renewcommand\color[2][]{}%
  }%
  \providecommand\transparent[1]{%
    \errmessage{(Inkscape) Transparency is used (non-zero) for the text in Inkscape, but the package 'transparent.sty' is not loaded}%
    \renewcommand\transparent[1]{}%
  }%
  \providecommand\rotatebox[2]{#2}%
  \ifx\svgwidth\undefined%
    \setlength{\unitlength}{233.80449219bp}%
    \ifx\svgscale\undefined%
      \relax%
    \else%
      \setlength{\unitlength}{\unitlength * \real{\svgscale}}%
    \fi%
  \else%
    \setlength{\unitlength}{\svgwidth}%
  \fi%
  \global\let\svgwidth\undefined%
  \global\let\svgscale\undefined%
  \makeatother%
  \begin{picture}(1,0.59968113)%
    \put(0,0){\includegraphics[width=\unitlength,page=1]{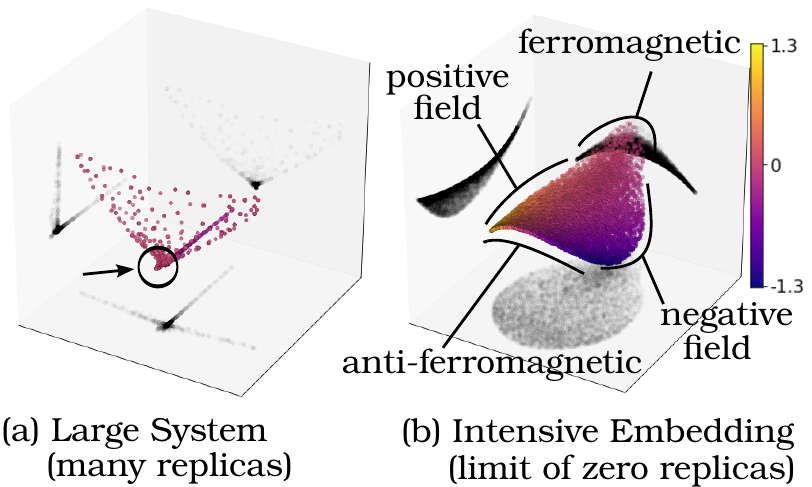}}%
  \end{picture}%
\endgroup%
    
    \caption{\textbf{Replicated Ising Model} illustrating the derivation of our intensive embedding. All points are coloured by magnetic field strength. (a)~Large dimensions are characterized by large system sizes; here we mimic a $128\times 128$ Ising model which is of dimension $2^{128^2}$. The orthogonality problem becomes manifest as all points are effectively orthogonal, producing a useless visualization with all points clustered in the cusp. (b)~Using replica theory, we tune the dimensionality of the system and consider the limit as the number of replicas goes to zero. In this way, we derive our intensive embedding. Note that the z-axis reflects a negative-squared distance, a property which allows violations of the triangle inequality and is discussed in the text.}
    \label{fig:replicas}
\end{figure}

We saw in Fig.~\ref{fig:hypersphere} that increasing the dimension of the data led to a saturation of the distance function Eq.~\ref{eq:spheredist}. This problem is referred to as the loss of relative contrast or the concentration of distances~\cite{beyer1999nearest}, and to overcome it requires a non-Euclidean distance function, discussed below. In the last section we saw the same saturation of distance could be achieved by adding replicas, increasing the embedding dimension. Figure~\ref{fig:replicas}{(a)} shows this process taken to an extreme: the model manifold of the $2\times 2$ Ising model with the number of replicas taken to infinity. All the points cluster together, obscuring the fact that the underlying manifold is two-dimensional. In order to cure the abundance of information which makes all points on the hypersphere equidistant, we seek an intensive distance, such as the distance per number of replicas observed. Next, because the limit of many replicas artificially leads to the same symptoms of the curse of dimensionality, we will consider the limit of zero replicas, a procedure which is often used in the study of spin glasses and disordered systems~\cite{Parisi79}. Figure~\ref{fig:replicas}(b) shows the result of this analysis, the intensive embedding, where the distance concentration has been cured, and the inherent two-dimensional structure of the Ising model has been recovered.

To find the intensive embedding, we must first find the distance between replicated models. The likelihood for $N$ replicas of a system is given by their product
\begin{eqnarray}
\label{eq:replicatedLike}
\Like{\{ \B{x_1},\dots,\B{x_N}\}\mid\B{\theta})}^{(N)} = \Like{\B{x_1}\mid\B{\theta}}\cdots\Like{\B{x_N}\mid\B{\theta}},
\end{eqnarray}
where the set $\{\B{x_1},\dots,\B{x_N}\}$ represents the observed data in the replicated systems. Writing the inner product or cosine angle between two distributions as
\begin{eqnarray}
\label{eq:CosAngle}
\left<\B{\theta_1};\B{\theta_2}\right> = \sqrt{\Like{\B{x}\mid\B{\theta_1}}}\cdot \sqrt{\Like{\B{x}\mid\B{\theta_2}}},
\end{eqnarray}
and using Eq.~\ref{eq:hellinger}, the distance per replica $d^2_N$ between two points on the model manifold is
\begin{eqnarray}
d_{N}^2(\B{\theta_1},\B{\theta_2}) = \frac{d^2(\B{\theta_1},\B{\theta_2})}{N}= -8\frac{\left<\B{\theta_1};\B{\theta_2}\right>^N-1}{N}.
\end{eqnarray}
We are now poised to define the intensive distance by taking the number of replicas to zero
\begin{eqnarray}
\label{eq:DI}
d_I^2(\B{\theta_1},\B{\theta_2}) = \lim_{N\rightarrow 0}d_N^2(\B{\theta_1},\B{\theta_2}) = -8\log\left<\B{\theta_1};\B{\theta_2}\right>.
\label{eq:intensivedist}
\end{eqnarray}
The last equality is achieved using the standard trick in replica theory, $(x^N - 1)/N \rightarrow \log x$ as $N\rightarrow\infty$, which is a basis trick used to solve challenging problems in statistical physics~\cite{Parisi79}. The trick is most evident using the identity $x^N=\exp(\log Nx)\approx 1+N\log x$. 
One can check that the intensive distance is isometric:
\begin{eqnarray}
d^2_I(\B{\theta},\B{\theta}+\delta\B{\theta}) = \delta\theta^\alpha\delta\theta^\beta g_{\alpha\beta} =\delta\theta^\alpha\delta\theta^\beta \mathcal{I}_{\alpha\beta},
\end{eqnarray}
where again $\mathcal{I}$ is the Fisher Information Metric in Eq.~\ref{eq:FIM}, so that we can be confident the intensive embedding distance preserves local structures. 

Importantly, the intensive distance does not satisfy the triangle inequality (and is thus non-Euclidean): the distance between points on the hypersphere can go to infinity, rather than lie constrained to the finite radius of the hypersphere embedding. Because of this, the intensive embedding can overcome the loss of relative contrast~\cite{beyer1999nearest} discussed at the beginning of this section. Distances in the intensive embedding maintain distinguishability in high dimensions, as illustrated in Fig.~\ref{fig:replicas}(b), wherein the two dimensional nature of the Ising model has been recovered. We hypothesize that this process, which cures the curse of dimensionality for models with too many samples, will also cure it for models with intrinsically high-dimensionality. The intensive distance obtained here is proportional to the Bhattacharyya distance~\cite{Bhattacharyya}. Considering the zero replica limit of the Hellinger divergence, we discovered a new way to derive the Bhattacharyya distance. The importance of this will be discussed further in the following section.

\subsection*{Connection to Least Squares}

Consider the concrete and canonical paradigm of models $f_i(\theta)$ with data points $x_i$ and additive white Gaussian noise, usually called a nonlinear least-squares model. The likelihood $\Like{\B{x}\mid\B{\theta}}$ is defined by
\begin{eqnarray}
-\log\Like{\B{x}\mid\B{\theta}} = \sum_i \frac{(f_i(\B{\theta})-x_i)^2}{2 \sigma_i^2} + \log \mathcal{Z}(\B{\theta}),
\end{eqnarray}
where $\mathcal{Z}$ sets the normalization. A straightforward evaluation of the intensive distance given by Eq.~\ref{eq:intensivedist} finds for the case of nonlinear least squares that
\begin{equation}
	d^2_I(\B{\theta_1},\B{\theta_2}) = \sum_i\frac{(f_i(\B{\theta_1}) - f_i(\B{\theta_2}))^2}{\sigma_i^2},
\end{equation}
so that the intensive distance is simply the variance-scaled Euclidean distance between model predictions.

\section*{Intensive Principal Component Analysis}

Classical Principal Component Analysis (PCA) takes a set of data examples and infers features which are linearly uncorrelated.~\cite{hotelling1933}. The features to be analyzed with PCA are compared via their Euclidean distance. Can we generalize this comparison to utilize our intensive embedding distance? Given a matrix of data examples $X\in\mathbb{R}^{m\times p}$ (with features along the rows), PCA first requires the mean-shifted matrix $M_{ij} = X_{ij} - \bar{X}_i = PX$, where $P_{ij} = \delta_{ij} - 1/p$ is the mean-shift projection matrix and $p$ is the number of sampled points. The covariance and its eigenvalue decomposition are then
\begin{equation}
	\mathrm{cov}(X,X) = \frac{1}{p} M^T M = X^T P P X = V \Sigma V^T,
\end{equation}
where the orthogonal columns of the matrix $V$ are the natural basis onto which the rows of $M$ are projected
\begin{equation}
    MV = (UDV^T)V = UD = U\sqrt{\Sigma},
    \label{eq:projection}
\end{equation}
where the columns of $U\sqrt{\Sigma}$ are called the principal components of the data $X$.

The principal components can also be obtained from the cross-covariance matrix, $M M^T$, since
\begin{eqnarray}
MM^T = P X X^T P = (U D V^T)(U D V^T)^T = U \Sigma U^T.
\end{eqnarray}
The eigenbasis $U$ of the cross-covariance is the natural basis for the components of the data, and the eigenbasis $V$ of the covariance is the natural basis of the data points. For us this flexibility is invaluable, as the cross-covariance is more natural for expressing the distances between distributions of different parameters.

Writing our data matrix as $X_{ij} = z_i(\theta_j)$ using Eq.~\ref{eq:firstembedding} for replicated systems, the cross-covariance is
\begin{eqnarray}
(MM^T)_{ij}^{(N)} &=& (P X X^T P)_{ij}\nonumber\\
&=& (\B{z}(\B{\theta_i}) - \bar{\B{z}})\cdot (\B{z}(\B{\theta_j}) - \bar{\B{z}}) \nonumber \\
&=& 4\left<\B{\theta}_i;\B{\theta}_j\right>^N + \frac{4}{p^2}\sum_{k,k'=1}^p \left<\B{\theta}_k;\B{\theta}_{k'}\right>^N \nonumber \\
&~& - \frac{4}{p} \sum_{k=1}^p \left(\left<\B{\theta}_{i};\B{\theta}_k\right>^N + \left<\B{\theta}_{j};\B{\theta}_k\right>^N \right),
\end{eqnarray}
where $\bar{z}$ is the average over all sampled parameters, and we used the definition of $z$ in Eqn.~\ref{eq:replicatedLike}. As with the intensive embedding, we can take the limit as the number of replicas goes to zero to find
\begin{eqnarray}
W_{ij} = \lim_{N\rightarrow 0} \frac{1}{N}(MM^T)_{ij}^{(N)}.
\end{eqnarray} 
Explicitly, the intensive cross-covariance matrix is
	\begin{eqnarray}
	\label{eq:W}
	W_{ij} &=& 4\log\left<\B{\theta_i};\B{\theta_j}\right> + \frac{4}{p^2}\sum_{k,k'=1}^p\log\left<\B{\theta_{k'}};\B{\theta_k}\right> \nonumber \\
    &~& - \frac{4}{p}\sum_{k=1}^p\left( \log\left<\B{\theta_{i}};\B{\theta_k}\right> + \log\left<\B{\theta_{j}};\B{\theta_k}\right> \right) \\
    &=& (P L P)_{ij}
	\end{eqnarray}
where $L_{i,j} = 4\log\left<\B{\theta_i};\B{\theta_j}\right>$ and $P$ is the same projection matrix as defined above. In taking the limit of zero replicas, the structure of the cross-covariance has transformed
\begin{eqnarray}
P XX^T P \xrightarrow[N\to 0]{} P L P \label{eq:PLP},
\end{eqnarray}
and thus the symmetric Wishart structure is lost. It is therefore possible to obtain negative eigenvalues in this decomposition, which give rise to imaginary components in the projections. Note the similarity between the form of this cross-covariance, and the double-centered distance matrix used in PCA and multidimensional scalding (MDS). This arises because both InPCA and PCA/MDS rely on mean-shifing the input data before finding an eigenbasis. Thus we view InPCA as a natural generalization of PCA to probability distributions, and MDS to non-Euclidean embeddings.

In summary, Intensive Principal Component Analysis (InPCA) is achieved by the following procedure:
\begin{enumerate}
	\item \textit{Compute the cross-covariance matrix from a set of probability samples:} Compute $W_{ij}$ as derived in Eq.~\ref{eq:W}.
	\item \textit{Compute the eigenvalue decomposition} $W = U\Sigma U^T$.
	\item \textit{Compute the coordinate projections,} $T=U\sqrt{\Sigma}$.
	\item \textit{Plot the projections} using the columns of $T$.
\end{enumerate}

\subsection*{Neural Network MNIST Digit Classifier}

\begin{figure}

\begingroup%
  \makeatletter%
  \providecommand\color[2][]{%
    \errmessage{(Inkscape) Color is used for the text in Inkscape, but the package 'color.sty' is not loaded}%
    \renewcommand\color[2][]{}%
  }%
  \providecommand\transparent[1]{%
    \errmessage{(Inkscape) Transparency is used (non-zero) for the text in Inkscape, but the package 'transparent.sty' is not loaded}%
    \renewcommand\transparent[1]{}%
  }%
  \providecommand\rotatebox[2]{#2}%
  \ifx\svgwidth\undefined%
    \setlength{\unitlength}{250.13134766bp}%
    \ifx\svgscale\undefined%
      \relax%
    \else%
      \setlength{\unitlength}{\unitlength * \real{\svgscale}}%
    \fi%
  \else%
    \setlength{\unitlength}{\svgwidth}%
  \fi%
  \global\let\svgwidth\undefined%
  \global\let\svgscale\undefined%
  \makeatother%
  \begin{picture}(1,0.76017312)%
    \put(0,0){\includegraphics[width=\unitlength,page=1]{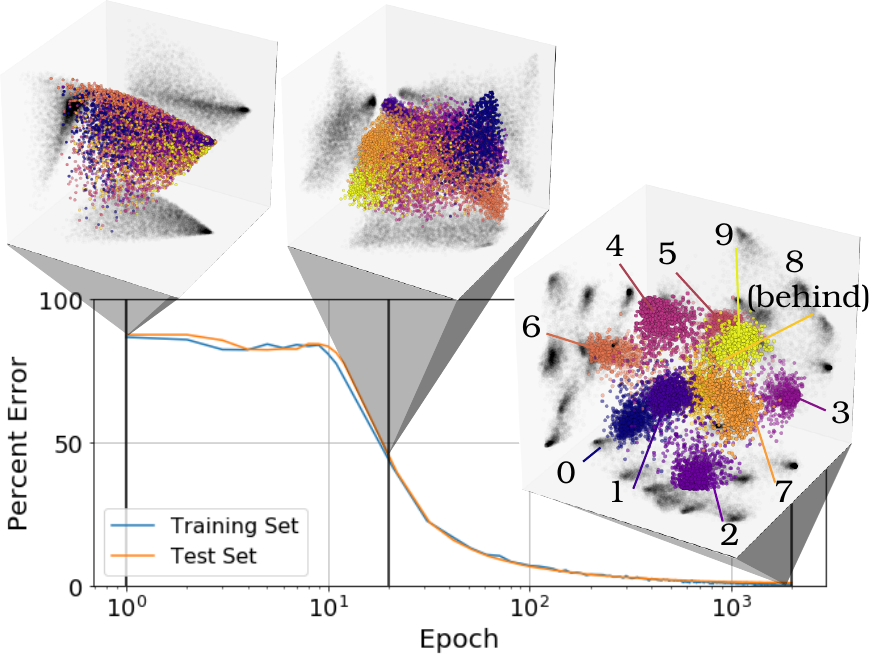}}%
  \end{picture}%
\endgroup%

    \caption{\textbf{Stages of training a convolutional neural network (CNN).} Each point in the above 3D projections represents one of 10,000 test image supplied to the CNN. At the first epoch, the neural network is untrained and so is unable to reliably classify images, with about a $90\%$ error rate -- an effect reflected in the cloud of points. As training progresses and error rate decreases, the cloud begins to cluster as shown by InPCA at the 20th epoch. Finally, when completely trained, the clustered regions are manifest at the 2000th epoch with ten clusters representing the ten digits.}
    \label{fig:manifolds}
\end{figure}

To demonstrate the utility of InPCA, we use it to visualize the training of a two layer convolution neural network (CNN), constructed using TensorFlow~\cite{TensorFlow},  trained on the MNIST data set of hand-written digits~\cite{MNIST}. A set of 55,000 images were used to train the network, which was then used to predict the likelihood that an additional set of 10,000 images are each classified as a specific digit between 0 and 9. We use softmax~\cite{Softmax} to calculate the probabilities from the category estimates supplied by the network. The CNN defines the likelihood $\Like{x\mid\theta}$ that some input image $\theta$ contains the image of a particular handwritten digit $x$. The InPCA projections of the CNN output in Fig.~\ref{fig:manifolds} visualizes the clustering learned by the CNN as a function of the number of learning epochs. The initialized network's model manifold shows no knowledge of the digits (colored dots), but as training commences, the manifold clearly separates digits into separate regions of its manifold (see supplemental animation). InPCA can be used as a fast, interpretable, and deterministic method for qualitatively evaluating what a neural network has learned.

\section*{Properties of the Intensive Embedding and InPCA}

The new space characterized by our intensive embedding has two weird properties: first it is formally one dimensional, yet there are multiple orthogonal directions upon which it can be projected, and second it is Minkowski-like, in that it has negative squared distances, violating the triangle inequality. We posit that, fundamentally, this second property is what allows InPCA to cure the orthogonality problem.

We begin with a discussion of the the one-dimensional nature of the embedding space. The embedding dimension is given by $D^N$ where $D$ is the original dimension of data $x$ and $N$ is the number of replicas. In the case of non-integer replicas the space becomes `fractional' in dimension, and in the limit of zero replicas ultimately goes to one. However, it is still possible to obtain projections themselves along the dominant components of this space, by leveraging the cross-covariance instead of the covariance, summarized in step 2 of our algorithm. Visualizations produced by InPCA are cross-sections of a space of the dimension equal to the number of sampled points of the model manifold $p$, instead of the dimension $D$ or $D^N$. 

In the limit of zero replicas in Eq.~\ref{eq:W}, the positive-definite, Wishart structure of the cross-covariance matrix is lost. It is therefore possible to have negative squared distances. The non-Euclidean nature of the embedding (flat, but Minkowski-like) does not suffer from the concentration of distances which plagues Euclidean measures in high dimensions, thus allowing the model manifold to be `unwound' from the $N$-sphere and for InPCA to produce useful, low-dimensional representations. 

Finally, the eigenvalues of InPCA correspond to the cross-sectional widths of the model manifold. We see this quite explicitly with the following example of a biased coin (specifically, in Fig.~\ref{fig:coinToss}{(b)}) where the eigenvalues extracted from InPCA map directly to the manifold widths measured along the direction of the corresponding InPCA eigenvector. Therefore, we see that InCA produces a hierarchy of directions, ordered by the global widths of the model manifold. Note that, as with classical PCA, this correspondence depends on how faithfully the model manifold was originally sampled, that is InPCA can only tell you about the structure of the manifold from observed points.

\subsection*{Biased Coins}

\begin{figure}

\begingroup%
  \makeatletter%
  \providecommand\color[2][]{%
    \errmessage{(Inkscape) Color is used for the text in Inkscape, but the package 'color.sty' is not loaded}%
    \renewcommand\color[2][]{}%
  }%
  \providecommand\transparent[1]{%
    \errmessage{(Inkscape) Transparency is used (non-zero) for the text in Inkscape, but the package 'transparent.sty' is not loaded}%
    \renewcommand\transparent[1]{}%
  }%
  \providecommand\rotatebox[2]{#2}%
  \ifx\svgwidth\undefined%
    \setlength{\unitlength}{250.28905106bp}%
    \ifx\svgscale\undefined%
      \relax%
    \else%
      \setlength{\unitlength}{\unitlength * \real{\svgscale}}%
    \fi%
  \else%
    \setlength{\unitlength}{\svgwidth}%
  \fi%
  \global\let\svgwidth\undefined%
  \global\let\svgscale\undefined%
  \makeatother%
  \begin{picture}(1,0.50623536)%
    \put(0,0){\includegraphics[width=\unitlength,page=1]{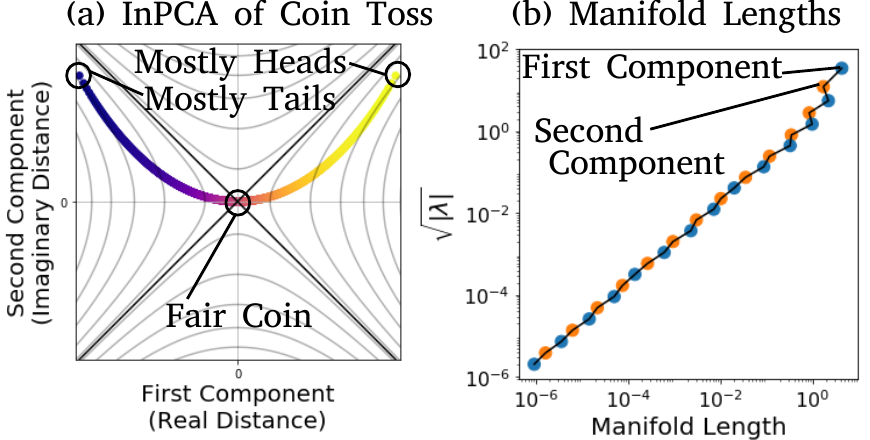}}%
  \end{picture}%
\endgroup%

    \caption{\textbf{InPCA visualization of biased coins}. (a)~The first two InPCA components correspond to the coin bias and variance, yet the first is real and the second is imaginary (the aspect ratio between axes is one). The contour lines representing constant distances from a fair coin and are hyperbolas: points can be a finite distance from a fair coin yet an infinite distance from each other. (b)~The ordered eigenvalues correspond to the manifold lengths, illustrating the hierarchical nature of the componnets extracted from InPCA.}
    \label{fig:coinToss}
\end{figure}

To illustrate the properties of InPCA, we use it to visualize a simple probabilistic model, that of a simple biased coin. A biased coin has one parameter, the odds ratio of heads to tails, and so forms a one-dimensional manifold.  Fig.~\ref{fig:coinToss}{(a)} shows the first two InPCA components for the manifold of biased coins, for 2,000 sampled points with probabilities uniformly spread between 0 and 1 (excluding the endpoints, since they are orthogonal and thus are infinitely far apart). The two extracted InPCA components correspond to the bias and variance of the coin, respectively. The hierarchy of components extracted from InPCA therefore correspond to known features of the model (i.e. they are meaningful).

The importance of the negative-squared distances is illustrated in Fig.~\ref{fig:coinToss}. The contour lines representing constant distances from a fair coin and are hyperbolas: points can be a finite distance from a fair coin yet an infinite distance from each other. As two
oppositely biased coins become increasingly biased, their distance from each other can go to infinity (because an outcome of a coin which always lands heads will never be the same as an outcome of a coin which always land tails) yet all remain a finite distance from a fair coin. Note that all points are in the left and right portions of the figure, representing net positive distances (the intensive pairwise distances are all positive).

\subsection*{Comparing with t-SNE and Diffusion Maps}

We compare our manifold learning technique to two standard methods, t-SNE and the diffusion maps by applying each to the six parameter \LCDM\ cosmological model predictions of the cosmic microwave background (CMB). The \LCDM\ predicts $\Like{x\mid\theta}$ where $x$ represents fluctuations in the CMB, and $\theta$ are the different cosmological parameters (i.e. it predicts the angular power spectrum of temperature and polarization anisotropies in sky maps of the CMB). Observations of the CMB from telescopes on satellites, balloons, and the ground provide thousands of independent measurements from large angular scales to a few arcminutes, that are use to fit model parameters. Here we only consider CMB observations from the 2015 Plank data release~\cite{PlanckData}. The \LCDM\ model we consider has six parameters, the Hubble constant ($H_0$) which we sampled in a range of 20 to 100~$km~s^{-1}~Mpc^{-1}$, the physical baryon density~($\Omega_b h^2$) and the physical cold dark matter density~($\Omega_c h^2$) both sampled from 0.0009 to 0.8, the primordial fluctuation amplitude~($A_s$) sampled from $10^{-11}$ to $10^{-8}$, the scalar spectral index~($\eta$) sampled from $0$ to 0.98, and the optical depth at reionization~($\tau$) sampled from 0.001 to 0.9.

\begin{figure}

\begingroup%
  \makeatletter%
  \providecommand\color[2][]{%
    \errmessage{(Inkscape) Color is used for the text in Inkscape, but the package 'color.sty' is not loaded}%
    \renewcommand\color[2][]{}%
  }%
  \providecommand\transparent[1]{%
    \errmessage{(Inkscape) Transparency is used (non-zero) for the text in Inkscape, but the package 'transparent.sty' is not loaded}%
    \renewcommand\transparent[1]{}%
  }%
  \providecommand\rotatebox[2]{#2}%
  \ifx\svgwidth\undefined%
    \setlength{\unitlength}{240bp}%
    \ifx\svgscale\undefined%
      \relax%
    \else%
      \setlength{\unitlength}{\unitlength * \real{\svgscale}}%
    \fi%
  \else%
    \setlength{\unitlength}{\svgwidth}%
  \fi%
  \global\let\svgwidth\undefined%
  \global\let\svgscale\undefined%
  \makeatother%
  \begin{picture}(1,0.71288879)%
    \put(0,0){\includegraphics[width=\unitlength]{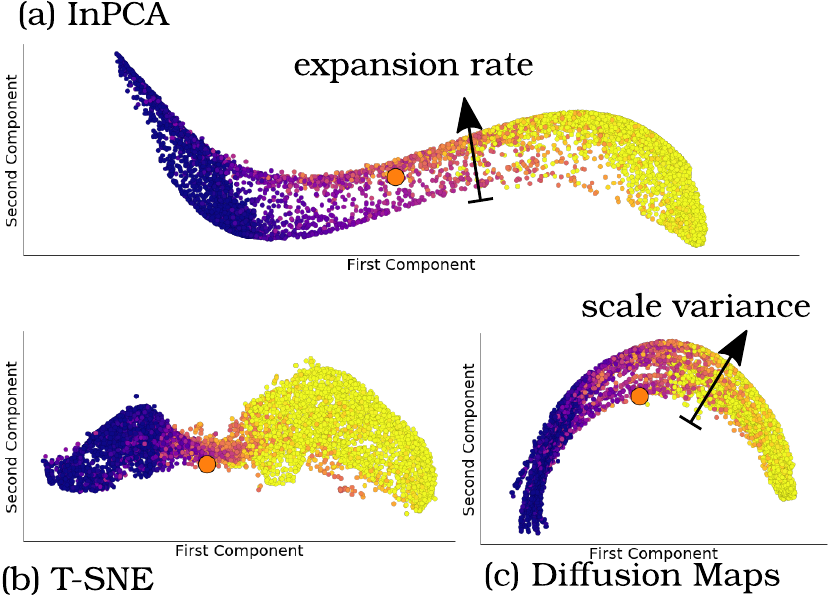}}%
  \end{picture}%
\endgroup%
    
    \caption{\textbf{Model manifold of the six parameter dark energy cold dark matter (\LCDM) cosmological model predictions of temperature and polarization power spectra in the CMB} using InPCA, t-SNE and the diffusion map. Axes reflect true aspect ratio from extracted components in all cases. Here the model manifold is colored by the primordial fluctuation amplitude, the most prominent feature in CMB data. (a)~InPCA extracts, as the first and second component, this amplitude term as well as the Hubble constant. These parameters control the two most dominant features in the Plank data, and so reflect a physically meaningful hierarchy of importance. In contrast, (b)~t-SNE only extract the amplitude term and (c)~the diffusion map extract the amplitude term and a different parameter, the scalar spectral index $\eta$, which reflects the scale variance of the density fluctuations in the early universe. In all plots, the orange point represents our universe, as represented by Planck~2015 data.}
    \label{fig:manifoldsCompare}
\end{figure}

To determine the likelihood functions, we use the CAMB software package to generate power spectra~\cite{Lewis:1999bs}. We perform a Monte Carlo sampling of 50,000 points around the best fit parameters provided by the 2015 Planck data release~\cite{PlanckData}, with sample weights based on the intensive distance to the best fit. 

In Fig.~\ref{fig:manifoldsCompare} we show the first two components of the manifold embedding for InPCA, t-SNE and diffusion maps. In order to apply t-SNE and the diffusion map to probabilistic data we must provide a distance. We therefore use our intensive distance, from Eq.~\ref{eq:DI}, for consistency and ease of comparison. In all three cases, the first component from each method is directly related to the primordial fluctuation amplitude $A_s$, which reflects the amplitude of density fluctuations in the early universe, and is the dominant feature in real data~\cite{PlanckData}. The second InPCA component predicts the Hubble constant, whereas the diffusion map predicts the scalar spectral index (a reflection of the size variance of primordial density fluctuations). In all cases, the projected components were plotted against the corresponding parameters to determine correlations, such as how one can see that $A_s$ corresponds with the first component in all three cases. Detailed plots and correlation coefficients for all three method are provided in the supplemental material.

Such stark differences between manifold learning methods are surprising, as all techniques aim to extract important features in the data distribution, i.e. important geometric features in the manifolds. Given the ranges of sampled parameters, one would expect the variation in the Hubble constant to relate in some way to one of the dominant components, as id does for InPCA. Figures illustrating the effect of different parameters are provided in the supplemental material, following results from~\cite{HuTutorial}.

There are two important differences between InPCA and other methods. First is that InPCA has no tunable parameters, and yields a geometric object defined entirely by the model distribution. For example, t-SNE embeddings rely on parameters such as the perplexity, a learning rate, and a random seed (yielding non-deterministic results), and the diffusion maps rely on a diffusion parameter and choice of diffusion operator, all of which must be manually optimized to obtain good results. Second, t-SNE and diffusion maps embed manifolds in Euclidean spaces in a way which aims to preserve local features. However, InPCA seeks to preserve both global and local features, by embedding manifolds in a non-Euclidean space.

\section*{Summary}

In this manuscript, we introduce an unsupervised manifold learning technique, InPCA, which captures low-dimensional features of general, probabilistic models with wide-ranging applicability. We consider replicas of a probabilistic system to tune its dimensionality and consider the limit of zero replicas, deriving an intensive embedding that ameliorates the canonical orthogonality problem. Our intensive embedding provides a natural, meaningful way to characterize a symmetric distance between probabilistic data and produces a simple, deterministic algorithm to visualize the resulting manifold.


\acknow{We thank Mark Transtrum for guidance on algorithms and for useful conversations. We thank Pankaj Mehta for pointing out the connection to MDS. KNQ was supported by a fellowship from the Natural Sciences and Engineering Research Council of Canada (NSERC), and JPS and KNQ were supported by the National Science Foundation through grant NSF DMR-1312160 and DMR-1719490. MDN was supported by NSF grant AST-1454881.}

\showacknow{} 

\bibliography{VisualizingProbabilisticModels}

\begin{thebibliography}{10}

\bibitem{Ferreira2003}
De~Oliveira MF, Levkowitz H (2003) From visual data exploration to visual data
  mining: a survey.
\newblock {\em IEEE Transactions on Visualization and Computer Graphics}
  9(3):378--394.

\bibitem{Liu2017}
Liu S, Maljovec D, Wang B, Bremer PT, Pascucci V (2017) Visualizing
  high-dimensional data: Advances in the past decade.
\newblock {\em IEEE Transactions on Visualization and Computer Graphics}
  23(3):1249--1268.

\bibitem{Lee2007}
Lee JA, Verleysen M (2007) {\em Nonlinear Dimensionality Reduction}.
\newblock (Springer, NY).

\bibitem{ArthurZimekErichSchubert2012}
Zimek A, Schubert E, Kriegel HP (2012) A survey on unsupervised outlier
  detection in high-dimensional numerical data.
\newblock {\em Statistical Analysis and Data Mining: The ASA Data Science
  Journal} 5(5):363--387.

\bibitem{murphy2012machine}
Murphy KP (2012) {\em Machine Learning: A Probabilistic Perspective}.
\newblock (The MIT Press).

\bibitem{kriegel2009clustering}
Kriegel HP, Kr\"{o}ger P, Zimek A (2009) Clustering high-dimensional data: A
  survey on subspace clustering, pattern-based clustering, and correlation
  clustering.
\newblock {\em ACM Trans. Knowl. Discov. Data} 3(1):1:1--1:58.

\bibitem{hotelling1933}
Hotelling H (1933) Analysis of a complex of statistical variables into
  principal components.
\newblock {\em Journal of educational psychology} 24(6):417.

\bibitem{Torgerson1952}
Torgerson WS (1952) Multidimensional scaling: I. theory and method.
\newblock {\em Psychometrika} 17(4):401--419.

\bibitem{VanDerMaaten2008}
Maaten Lvd, Hinton G (2008) Visualizing data using t-sne.
\newblock {\em Journal of machine learning research} 9(Nov):2579--2605.

\bibitem{Coifman2005}
Coifman RR, et~al. (2005) Geometric diffusions as a tool for harmonic analysis
  and structure definition of data: Diffusion maps.
\newblock {\em Proceedings of the National Academy of Sciences}
  102(21):7426--7431.

\bibitem{mcinnes2018umap}
McInnes L, Healy J, Melville J (2018) Umap: Uniform manifold approximation and
  projection for dimension reduction.
\newblock {\em arXiv preprint arXiv:1802.03426}.

\bibitem{MezardPV87}
M{\'e}zard M, Parisi G, Virasoro M (1986) {\em Spin Glass Theory and Beyond}.
\newblock (WORLD SCIENTIFIC).

\bibitem{MachtaCTS13}
Machta BB, Chachra R, Transtrum MK, Sethna JP (2013) Parameter space
  compression underlies emergent theories and predictive models.
\newblock {\em Science} 342(6158):604--607.

\bibitem{Transtrum2014}
Transtrum MK, Qiu P (2014) Model reduction by manifold boundaries.
\newblock {\em Phys. Rev. Lett.} 113(9):098701.

\bibitem{TranstrumSloppyReview}
Transtrum MK, et~al. (2015) Perspective: Sloppiness and emergent theories in
  physics, biology, and beyond.
\newblock {\em The Journal of Chemical Physics} 143(1):010901.

\bibitem{hellinger}
Hellinger E (1909) {Neue Begr\"undung der Theorie quadratischer Formen von
  unendlichvielen Ver\"anderlichen.}
\newblock {\em {J. Reine Angew. Math.}} 136:210--271.

\bibitem{gromov2013search}
Gromov M (2013) In a search for a structure, part 1: On entropy.

\bibitem{Amari}
Amari S, Nagaoka H (2000) {\em Translations of Mathematical Monographs: Methods
  of Information Geometry}.
\newblock (Oxford University Press) Vol.{} 191.

\bibitem{beyer1999nearest}
Beyer K, Goldstein J, Ramakrishnan R, Shaft U (1999) When is ``nearest
  neighbor'' meaningful? in {\em Database Theory --- ICDT'99}.
\newblock (Springer Berlin Heidelberg, Berlin, Heidelberg), pp. 217--235.

\bibitem{Parisi79}
Parisi G (1979) Infinite number of order parameters for spin-glasses.
\newblock {\em Phys. Rev. Lett.} 43(23):1754--1756.

\bibitem{Bhattacharyya}
Bhattacharyya A (1946) On a measure of divergence between two multinomial
  populations.
\newblock {\em Sankhy\={a}: The Indian Journal of Statistics (1933-1960)}
  7(4):401--406.

\bibitem{TensorFlow}
Abadi M, , et~al. (2015) {TensorFlow}: Large-scale machine learning on
  heterogeneous systems.
\newblock Software available from \url{https://tensorflow.org}.

\bibitem{MNIST}
LeCun Y, Cortes C, Burges CJ (2018) Mnist database, (Courant Institute, Google
  Lab, Microsoft Research), Technical report.

\bibitem{Softmax}
Bishop CM (2006) {\em Pattern Recognition and Machine Learning}.
\newblock (Springer, NY).

\bibitem{PlanckData}
{Planck~Collaboration} (2016) Planck 2015 results - i. overview of products and
  scientific results.
\newblock {\em A\&A} 594:A1.

\bibitem{Lewis:1999bs}
Lewis A, Challinor A, Lasenby A (2000) Efficient computation of cosmic
  microwave background anisotropies in closed friedmann-robertson-walker
  models.
\newblock {\em The Astrophysical Journal} 538(2):473--476.

\bibitem{HuTutorial}
Hu W (2001) Cmb tutorials.

\end{thebibliography}

\end{document}